\documentclass[]{tMOP2e}

\usepackage{color}

\newcommand{\be}{\begin{equation}}
\newcommand{\ee}{\end{equation}}
\newcommand{\bea}{\begin{eqnarray}}
\newcommand{\eea}{\end{eqnarray}}

\newcommand{\e}{\mathrm{e}}

\newcommand{\ket}[1]{{|#1\rangle}}

\newcommand{\vect}[1]{{\boldsymbol #1}}

\newcommand{\su}{\ket{\hspace{-2pt}\uparrow}}
\newcommand{\sd}{\ket{\hspace{-2pt}\downarrow}}

\newcommand{\fud}[0]{\frac{1}{2}}

\begin{document}

\doi{10.1080/0950034YYxxxxxxxx}
 \issn{1362-3044}
\issnp{0950-0340} \jvol{00} \jnum{00} \jyear{2010} \jmonth{10 January}

\markboth{Taylor \& Francis and I.T. Consultant}{Journal of Modern Optics}

\title{Prospective applications of optical quantum memories}
%

\author{
F\'elix Bussi\`eres$^{a}$$^{\ast}$\thanks{$^\ast$Corresponding author. Email: felix.bussieres@unige.ch}, Nicolas Sangouard$^{a}$, Mikael Afzelius$^{a}$, Hugues de Riedmatten$^{b,c}$, Christoph Simon$^{d}$ and Wolfgang Tittel$^{d}$\\ \vspace{6pt}  
$^{a}${\em{Group of Applied Physics, University of Geneva, CH-1211 Geneva 4, Switzerland}};
$^{b}${\em{ICFO-Institut de Ciencies Fotoniques, Av. Carl Friedrich Gauss 3, 08860 Castelldefels (Barcelona), Spain}};
$^{c}${\em{ICREA-Instituci\'o Catalana de Recerca i Estudis Avan\c{c}ats, 08015 Barcelona, Spain}};
$^{d}${\em{Institute for Quantum Science and Technology, and Department of Physics \& Astronomy, University of Calgary, Canada}}
\\\vspace{6pt}\received{Received ...; final version received ...} 
}

\maketitle

\begin{abstract}
An optical quantum memory can be broadly defined as a system capable of storing a useful quantum state through interaction with light at optical frequencies. During the last decade, intense research was devoted to their development, mostly with the aim of fulfilling the requirements of their first two applications, namely quantum repeaters and linear-optical quantum computation. A better understanding of those requirements then motivated several different experimental approaches. Along the way, other exciting applications emerged, such as as quantum metrology, single-photon detection, tests of the foundations of quantum physics, device-independent quantum information processing and nonlinear processing of quantum information. Here we review several prospective applications of optical quantum memories with a focus on recent experimental achievements pertaining to these applications.   
This review highlights that optical quantum memories have become essential for the development of optical quantum information processing.
\smallskip

\begin{keywords} Quantum memories, quantum information processing, light-atom interactions, atomic physics.
\end{keywords}
\end{abstract}


 

\section{Introduction}
Optical quantum information processing (QIP) is at the core of several applications such as quantum key distribution, computing and metrology. Harnessing the full potential of these applications is an important challenge with rewarding consequences, both for technological innovations and to understand the power offered by quantum physics itself. The development of these technologies relies on the availability of application-specific optical quantum memories. Indeed, it was recognized early on that the ability to store quantum states of light, and to retrieve them on-demand, was necessary to extend the transmission distance beyond the limit imposed by loss in the quantum channels~\cite{Briegel1998a}. This resource was also known to be essential to implement linear-optical quantum computing~\cite{Knill2001a,Kok2007a}, or, as a matter of fact, in any application requiring to synchronize several independent and probabilistic processes such as the creation of photon pairs through spontaneous parametric downconversion. This triggered intense experimental efforts, whose proliferation eventually lead to the writing of several reviews with varying focuses, such as quantum storage in atomic ensembles and trapped ions~\cite{Lukin2003a,Kimble2008a,Duan2010a}, optical quantum memories~\cite{Lvovsky2009a}, light-matter interactions~\cite{Hammerer2010a}, photon-echo quantum memories based on solid-state systems~\cite{Tittel2010a}, quantum memories in the European integrated project ``QAP''~\cite{Simon2010a}, quantum repeaters based on atomic ensembles and linear optics~\cite{Sangouard2011a}, quantum memories and quantum error correction~\cite{Wooton2012a}. 

The initial motivations to build optical quantum memories have now grown towards building a useful ingredient, or sometimes a crucial one, in essentially all the other applications of optical QIP. This paper is an application-oriented review of some recent experimental achievements making use of optical quantum memories to perform a useful, quantum-enhanced task. Its pertinence emerges from the rapid pace at which research into optical quantum memories evolves. With the intent of keeping this text succinct, we review almost exclusively experimental work published between 2010 and now, as well as some theoretical proposals that are directly relevant to the applications we are considering. Note that the different types of quantum memories that are presented as being suitable for a given application are not necessarily restricted to this application only. This review is therefore not extensive. Nevertheless, it is our hope that it will faithfully depict the state-of-the art of optical quantum memories, and stimulate further research in this exciting field.

Before going any further, we now define in more precise terms what we mean by \emph{optical quantum memory}. A quantum memory is, in broad terms, a system that can store a quantum state to accomplish a certain task. An ``optical'' quantum memory is one whose state can be prepared and/or manipulated using light in the visible to near-infrared range. This comprehensive definition was intentionally chosen to allow us to focus on several applications that we review here, namely quantum repeaters (section~\ref{section-repeaters}), linear-optical quantum computation (section~\ref{section-LOQC}), quantum metrology and magnetometry (section~\ref{section-metrology}), single-photon detection (section~\ref{section-detection}) and foundations of quantum mechanics (section~\ref{section-foundations}). We conclude with an outlook focusing on prospective quantum memories with built-in nonlinear processing capabilities (section~\ref{section-outlook}).

\section{Quantum repeaters} \label{section-repeaters}
The possibility to reversibly transfer quantum states between light and matter significantly benefits the field of long-distance quantum communication. In this section, we will discuss such a quantum memory for light in the context of a quantum repeater, which was introduced by Briegel~\textit{et al.} in 1998 \cite{Briegel1998a} to overcome decoherence during the transmission of quantum states. It was rapidly noted that the same idea can also be used to overcome limitations due to photon loss, which currently determines the maximum reach of long-distance quantum communications. As a matter of fact, most experimental groups (including our groups) working on the development of quantum repeaters currently only focus on the question how to avoid limitations due to loss, leaving more complicated setups that also allow dealing with decoherence for the future. In the next paragraph we will briefly review a simplified version of a quantum repeater that is inspired by~\cite{Duan2001a,Simon2007a}. Note that it does not include purification, i.e.~the aim is to speedup the rate at which entanglement can be distributed beyond the rate of direct transmission limited by loss only. 

A quantum repeater is the application of quantum memories that so far received most attention. In particular, quantum repeater architectures based on atomic ensembles and linear optics to perform entanglement swapping operations, have been intensively studied. Many architectures of this kind have been compared and the requirements on the memories have been clearly established, as reviewed in~\cite{Sangouard2011a}. Alternative appealing architectures using photon-photon gates~\cite{Munro2012a} or dissipation-driven entanglement creation and distillation~\cite{Vollbrecht2011a} have been proposed. However, we concentrate here only on the requirements for implementing quantum repeaters based on atomic ensembles and linear optics. They are today one of the most important motivation for building efficient and long-lived quantum memories.  

\subsection{Long-distance quantum communication with quantum repeaters}

All quantum communication protocols rely on, or can be implemented using, close-to-maximally entangled states. For simplicity, let us consider only protocols with two legitimate users, Alice and Bob, and assume they each hold one photon of a pair prepared in a maximally entangled two-qubit ``Bell'' state, which are $\ket{\Psi^\pm}= \frac{1}{\sqrt{2}}\left (\ket{0}_A\ket{1}_B\pm \ket{1}_A\ket{0}_B\right )$, $\ket{\Phi^\pm}= \frac{1}{\sqrt{2}}\left (\ket{0}_A\ket{0}_B\pm \ket{1}_A\ket{1}_B\right )$, where $\ket{0}$ and $\ket{1}$ form a single-qubit basis. 
Given several copies of the same Bell state, Alice and Bob can perform quantum key distribution by doing single qubit measurements on their respective photons and thereby establish a shared secret key from the quantum correlations, as proposed by Ekert in 1991~\cite{Ekert1991a,Bennett92a}. Furthermore, an unknown quantum state encoded into a third photon can be teleported from Alice to Bob using shared entanglement as a resource, as pointed out by Bennett \emph{et al.}~in 1993 \cite{Bennett1993a}. 

\begin{figure}[!t]
\begin{center}
\includegraphics[width=0.8\columnwidth]{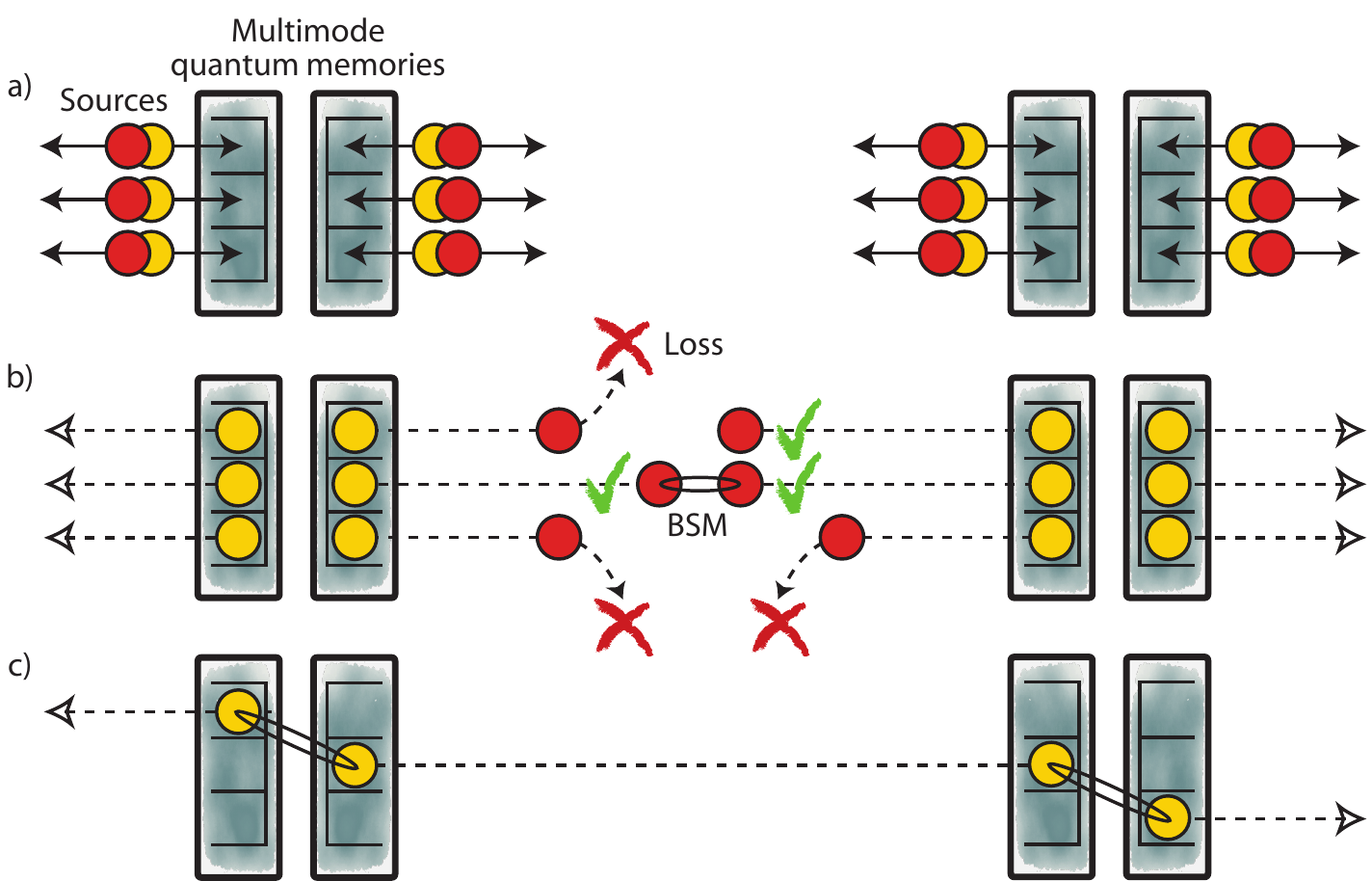}
\caption{Simplified version of a quantum repeater. The sequence shows how heralded entanglement can be created between two adjacent nodes constituting an \emph{elementary link} of a quantum repeater. Each node is composed of $2N$ probabilistic sources of entangled photon pairs and two multimode quantum memories capable of storing $N$ modes each ($N=3$ on the figure). a) All sources are triggered simultaneously and one photon from each pair is stored in the quantum memory (in practice, the creation of an entangled pair is a probabilistic process). b) The other photons are sent towards the adjacent node, where they meet halfway. The photons that were not lost are subjected to a Bell state measurement (BSM) represented by an oval. The entanglement carried between two photons (represented by a dashed line) is then swapped to the photons stored in the memories. The BSM heralds the entanglement between two quantum memories. Entanglement is then stored until the neighbouring memories have also successfully stored entanglement. c) Once neighbouring quantum memories have both stored one excitation, photons are released and subjected to a BSM that swaps the entanglement to nodes that are further away. This process is repeated until entanglement is generated over the desired distance.  This essence of this scheme can also be realized with quantum memories that directly emit a single photon that is entangled with an internal degree of freedom of the memory~\cite{Sangouard2011a}. Similarly, the correlation between the single photon presence and the excitation state of the memory (in the form of ``single-photon entanglement'') can also be exploited directly~\cite{Duan2001a,Simon2007a,Sangouard2011a}. 
}%
\label{fig:QRep}
\end{center}
\end{figure}

Unfortunately, the distribution of entanglement over long distances suffers from photon loss and decoherence during transmission, and quantum communication based on the direct transmission of entanglement is limited to a distance of a few hundreds of kilometres at most. Surprisingly, it is possible to overcome these obstacles and establish entanglement across arbitrarily long distances using a quantum repeater. As depicted in Fig.~\ref{fig:QRep}, the basic idea is to divide a long quantum channel into shorter segments and to distribute entanglement in a heralded fashion between end nodes of these segments. Next, entanglement is extended over the entire link by means of entanglement swapping~\cite{Zukowski1993a}. Quantum memories are essential in the repeater protocol as the initial distribution of entanglement is of probabilistic nature. Quantum memories allow for the storage of entanglement in a given segment until  entanglement has also been established in the adjacent sections. This makes the different segments independent and removes the necessity for all probabilistic steps to succeed at the same time.


\subsection{Performance criteria for quantum repeaters}
In the following we will briefly discuss the required performance criteria in view of the development of a quantum repeater. 
\begin{itemize}
\item \textbf{Efficiency} -- The required storage efficiency essentially depends on all the other parameters. It can be estimated by numerical simulations, which all show that it should be close to one, \emph{e.g.} 90\% in~\cite{Sangouard2011a}. We note, however, that the entanglement distribution rate increases exponentially fast with the efficiency of the memories. For example, the analysis of several protocols show that a 1\%~increase in storage efficiency can increase the entanglement distribution rate by 7 to 18\%~\cite{Sangouard2011a}.
\item \textbf{Storage time} -- The required storage time is generally determined by the time it takes to establish entanglement across the entire repeater link, e.g. a few seconds across a 1000-km long fibre-link \cite{Sangouard2011a}. However, as shown in \cite{Munro2010a}, it is possible to reduce this time to the time it takes to establish entanglement across an elementary link, provided that all steps in a quantum repeater succeed deterministically. This can in principle be achieved by implementing each probabilistic steps many times in parallel fashion, thereby trading the need for long storage times against resources.
\item \textbf{Read-out on demand} -- An important aspect of quantum memories is to synchronize the release of photons for subsequent Bell state measurement between neighbouring quantum memories (Fig.~\ref{fig:QRep}-c). This requires recall on demand, i.e.~the possibility to determine the storage time after storage has occurred. Most optical quantum memories based on light-atom interaction (the ones considered in this review) naturally allow for this on-demand recall due to the need to actively trigger re-emission. 
\item \textbf{Fidelity} -- An optical memory is said to operate in the quantum regime if the fidelity of the retrieved single-photon quantum state $\rho_{out}$ (averaged over all possible input states $\rho_{in}$) exceeds the maximum fidelity of a reference process that does not involve any storage. One natural choice is to compare against the maximum fidelity of a \emph{prepare-and-measure} strategy, which consists in measuring the photon directly and preparing another photon whose state is as close as possible to the input state, based on the measurement result.  Assuming the quantum information is encoded in a single qubit, the largest achievable fidelity of such a strategy is $F = \rm{Tr}(\rho_{in}\rho_{out}) =2/3$~\cite{Massar1995a}. Another relevant figure is the minimum storage fidelity required to preserve the nonlocal nature of a stored-and-retrieved photon belonging to an entangled pair of qubits. In this case, the fidelity should be at least $85.4\%$. Note that overcoming theses thresholds alone is not sufficient to beat direct transmission with a quantum repeater, since unfaithful storage has to be compensated for by entanglement purification or error correction, which reduces the entanglement distribution rate.
\item\textbf{Appropriate wavelength} -- For long-distance communication over optical fibres, it is important that the photon travelling from the source to the Bell state measurement (see Fig.~\ref{fig:QRep}) is at a wavelength situated in the low-loss window of optical fibers (around 1550~nm). This can be very hard to realize in the DLCZ scheme~\cite{Duan2001a} at it puts stringent constraints on the material's atomic level structure, which determines the wavelength of the photon (see sections~\ref{section:coldatoms} and \ref{section:singlequantumsystems}).  The same applies to a single quantum system emitting a photon entangled with one of its internal degree of freedom (see section~\ref{section:singlequantumsystems} below).
This problem can be tackled using frequency conversion in nonlinear waveguides~(see for example \cite{Curtz2010a,Zaske2012a,DeGreve2012a}) or in atomic ensembles~\cite{Radnaev2010a}. Another solution is to avoid this problem from the start by interfacing sources of entangled photons, where one photon of each pair is at a telecommunication wavelength, with the storage and recall of the other photon in an optical quantum memory~\cite{Simon2007a}. 
\item \textbf{Multi-mode capacity} -- The time needed to attempt the creation of heralded entanglement between two remote quantum memories of an elementary link (see Fig.~\ref{fig:QRep}) is limited by the time for both photons to meet at midway, plus the time for both memories to receive the confirmation that the Bell state measurement succeeded (or not). This can severely limit the rate of entanglement creation if the memories can store at most one photon at the time. As shown in~\cite{Simon2007a}, the use of \emph{multi-mode} memories that can simultaneously store $N$ photons (that are part of $N$ entangled pairs) allows to time-multiplex the procedure. In turn, this yields an $N$-fold increase in success probability for entanglement creation per round-trip time, and thereby decreases the time to establish entanglement between end notes by the same factor. 
\item \textbf{Robustness and ease-of-use} -- Finally, even though not essential for proof-of-principle experiments, it is obvious that the deployment of quantum repeaters over real-world fibre links will only be possible if the underpinning technology, in particular quantum memories, is sufficiently robust and simple to use. The recent development of quantum memory based on atomic vapour cells operated above room temperature~\cite{Julsgaard2004a,Hosseini2011a,Reim2011a} and rare-earth-ion doped crystals (operated at 3K using push-button close-cycle coolers) \cite{Riedmatten2008a}, including crystals featuring waveguides \cite{Saglamyurek2012a}, are promising towards this end. 
\end{itemize}

In part triggered by the improved understanding of necessary properties, and in part causing it, impressive progress in the storage and recall of quantum states in atomic ensembles and individual absorbers has been achieved over the last few years. This includes storage efficiencies of up to 87\% \cite{Hedges2010a,Hosseini2011b}, storage over 5 GHz bandwidth~\cite{Saglamyurek2012a}, simultaneous storage of several temporal modes \cite{Usmani2010a,Bonarota2011a}, recall fidelities exceeding 99\%~\cite{Zhou2012a}, the combination of high efficiency (73\%) and long storage time (3~ms)~\cite{Bao2012a}, and storage and recall of members of entangled photon pairs \cite{Zhang2011a,Clausen2011a,Saglamyurek2011a}. Yet, the development of quantum memory suitable for a quantum repeater remains a significant challenge. While all required properties have meanwhile been demonstrated, this was achieved using different storage materials and protocols. The following section details some of the recent progress in the development of optical quantum memories for quantum repeaters.

\subsection{Quantum memories for quantum repeaters}  \label{section:qmrepeaters}
Several systems have been used to implement
quantum memories, including laser-cooled atomic ensembles, hot
atomic vapours, single atoms in high-finesse cavities and atomic
ensembles in rare-earth-ion doped crystals. During the last decade, the field progressed tremendously. 
Here, we will concentrate only on some recent developments.

\subsubsection{Cold atomic ensembles} \label{section:coldatoms}
\begin{figure}[t]
\begin{center}
\includegraphics[scale=0.75]{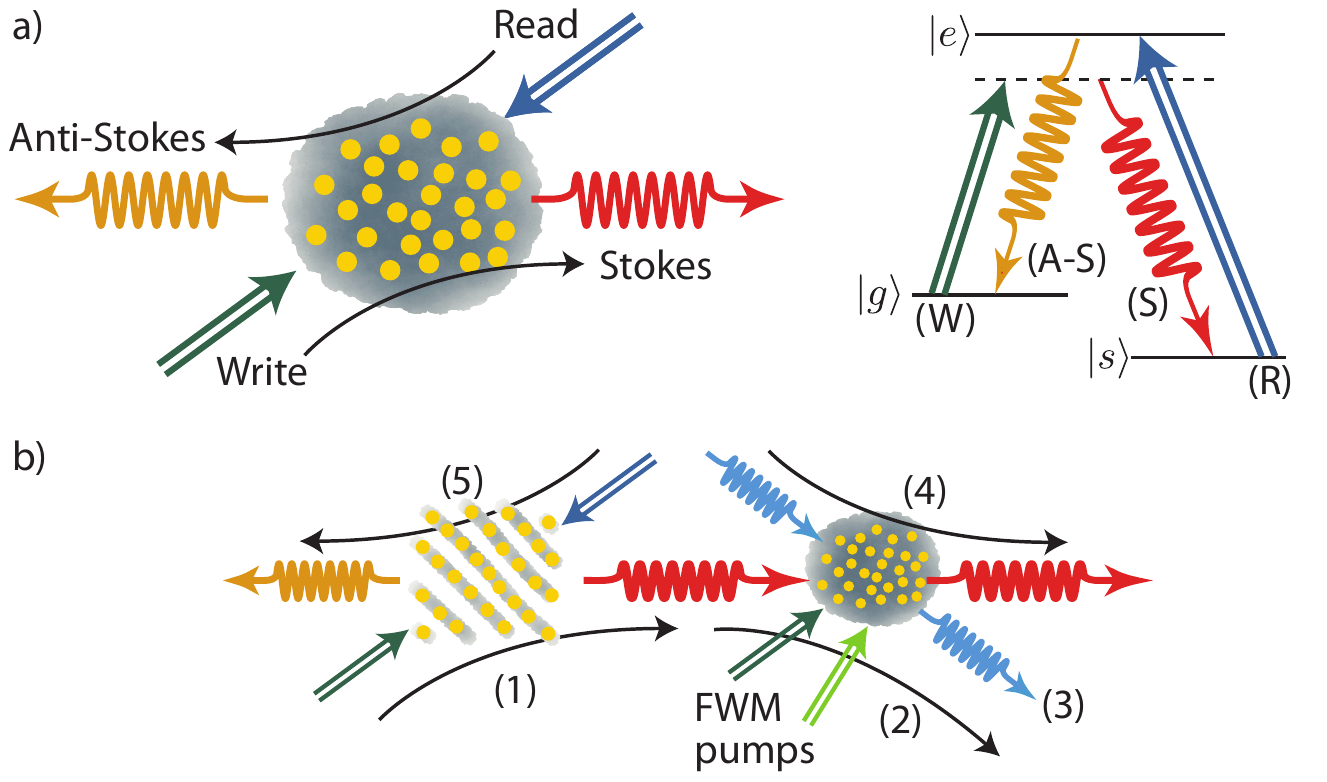}
\caption{DLCZ scheme. a) The DLCZ scheme consists of a cold atomic ensemble whose population is initially prepared in the ground state $\ket{g}$. An off-resonant write pulse of strong light (W) creates a Raman-scattered Stokes photon (S), whose detection heralds a collective excitation stored as a spin wave between levels $\ket{g}$ and $\ket{s}$. A read pulse of light (R), resonant on the $s$-$e$ transition, then maps the spin wave into an anti-Stokes photon (A-S) whose momentum obeys a strict phase-matching condition, and can thus be efficiently collected in an optical fiber~\cite{Duan2001a}. b) The experiment of Ref.~\cite{Radnaev2010a} starts with an atomic ensemble cooled in an optical lattice. A write pulse first creates a stokes photon at 795~nm (1) that is sent into another cold atomic ensemble to be downconverted (2) to a photon at 1367~nm (3) through a four-wave-mixing (FWM) process with non-degenerate pumps. This photon travels 100~m of optical fibre to be sent into the same ensemble, where it is converted back to 795~nm (4) to be detected. This detection heralds the presence of a stored excitation in the ensemble trapped in the optical lattice, which can be then converted to an anti-stokes photon~(5).}
\label{fig:dlcz}
\end{center}
\end{figure}

Laser-cooled atomic ensembles are currently one of the most advanced system for
light-matter interaction at the quantum level. One very important
class of quantum memories is based on the creation, storage and
retrieval of single collective spin excitations following the influential 
proposal of Duan, Lukin, Cirac and Zoller~\cite{Duan2001a}; see Fig.~\ref{fig:dlcz}-a for a description of this scheme. The realization of the building blocks of this scheme has been the subject of several founding experiments using cold atoms in magneto-optical traps (see~\cite{Kimble2008a,Sangouard2011a} for reviews). 
The most recent experiments based on this so-called DLCZ scheme elegantly illustrate its potential for creating high-efficiency quantum memories with 
long storage times. 

In 2010, a significant advance was reported through the demonstration of a DLCZ quantum memory with a storage time of 100~ms, 
which is the longest storage time reported in the
quantum regime~\cite{Radnaev2010a}. The authors used a cold atomic
Rubidium 87 ensemble loaded in an optical lattice to prevent decoherence
due to atomic motion; see Fig.~\ref{fig:dlcz}-c. An interesting aspect of that experiment is
that the Stokes photons at 795~nm emitted during the write process
were subsequently converted to photons at telecommunication
wavelengths (1367~nm) using a four-wave mixing process in a
separate cold atomic ensemble with high absorption. A
conversion efficiency of 50\% was achieved with an optical
depth of 150. The reverse process was used
to transfer the telecom photon back to a wavelength of 795~nm, where
it could be efficiently detected using a standard silicon avalanche photodiode used as a single-photon counter. Using a similar setup, an experiment demonstrating entanglement between a light-shift-compensated spin wave and a
photon at telecommunication wavelength was reported~\cite{Dudin2010a}. A violation of a Bell
inequality was observed after a storage time of 10~ms, confirming
that the entanglement between the quantum memory and the emitted photon is preserved during the frequency conversion of the Stokes photon. Recent work shows that the storage time of classical light can be extended to 16~s with the help of dynamical decoupling techniques~\cite{Dudin2013a}, where the main limitation for the storage time is the lifetime of the optical trap. 
While the storage time was extremely long in these experiment, the retrieval
efficiency was limited to around 20\% or less due to the limited number
of atoms that could be trapped in the optical lattice. 

Long storage time and high-efficiency were combined in a single setup in 2012~\cite{Bao2012a}. By placing a cold Rb atomic ensemble inside a ring cavity to enhance the write and read processes through the Purcell, the authors could observe a quantum memory with an intrinsic spin-wave-to-photon
conversion efficiency of 73\% together with a storage lifetime of 3.2~ms. 

Another interesting development for quantum networks based on the DLCZ scheme has
been reported in \cite{Choi2010a} where the authors demonstrate heralded entanglement between four atomic ensembles located in
the same atom trap.

Important progress has also been reported recently for quantum
storage based on electro-magnetically induced transparency. A
single photon from a polarization entangled photon pair created by
cavity-enhanced spontaneous downconversion has been stored and
retrieved in a cold Rb atomic ensemble \cite{Zhang2011a},
demonstrating polarization entanglement between a stored spin
excitation and a photon. The storage and retrieval efficiency  at
zero delay was about 10\% and the non-classical correlations
between the photon and the stored excitation remained non-classical for storage time of up to a few $\mu$s. In addition, the
violation of a Bell inequality was demonstrated for a storage time
of 200~ns. This experiment has been extended recently to the
storage of the two entangled photons from the pair (i.e. 4 modes
storage) using holographic storage in a cold atomic ensemble, 
with an efficiency reaching 16\%~\cite{Dai2012a}.

The storage and retrieval efficiency using electromagnetically-induced transparency (EIT) has been studied in
the classical regime in Ref.~\cite{Zhang2011b}. The authors used a
dense cold atomic $^{85}$Rb ensemble with variable optical depth up to~250. They found that the maximal efficiency of the storage was
limited to around 50\% due to a detrimental four-wave-mixing
effect involving the two ground states. This study confirms what
had been previously measured in hot atomic vapours~\cite{Novikova2007a}. 
However, a very recent study involving EIT storage of bright pulses in cold $^{87}$Rb ensemble
showed significantly higher storage and retrieval efficiencies up
to 78\%~\cite{Chen2013a}. High-efficiency EIT storage of single photons
generated using a cold atomic ensemble has also been recently
demonstrated in a cold atomic ensemble~\cite{Zhou2012b}.

Overall, the recent progress of optical quantum memories based on cold atomic ensembles is impressive, and several of the quantum repeater performance criteria are within reach. The main limitations are their intrinsically narrow optical bandwidth and their limited multimodal capacity. Nonetheless, the Raman scheme described in the following could provide a solution for the bandwidth limitation and approaches to increase the multimode capacity has been proposed~\cite{Simon2010b} but are yet to be implemented.

\subsubsection{Atomic vapours} \label{section:atomicvapors}
\begin{figure}[!t]
\begin{center}
\includegraphics[scale=0.75]{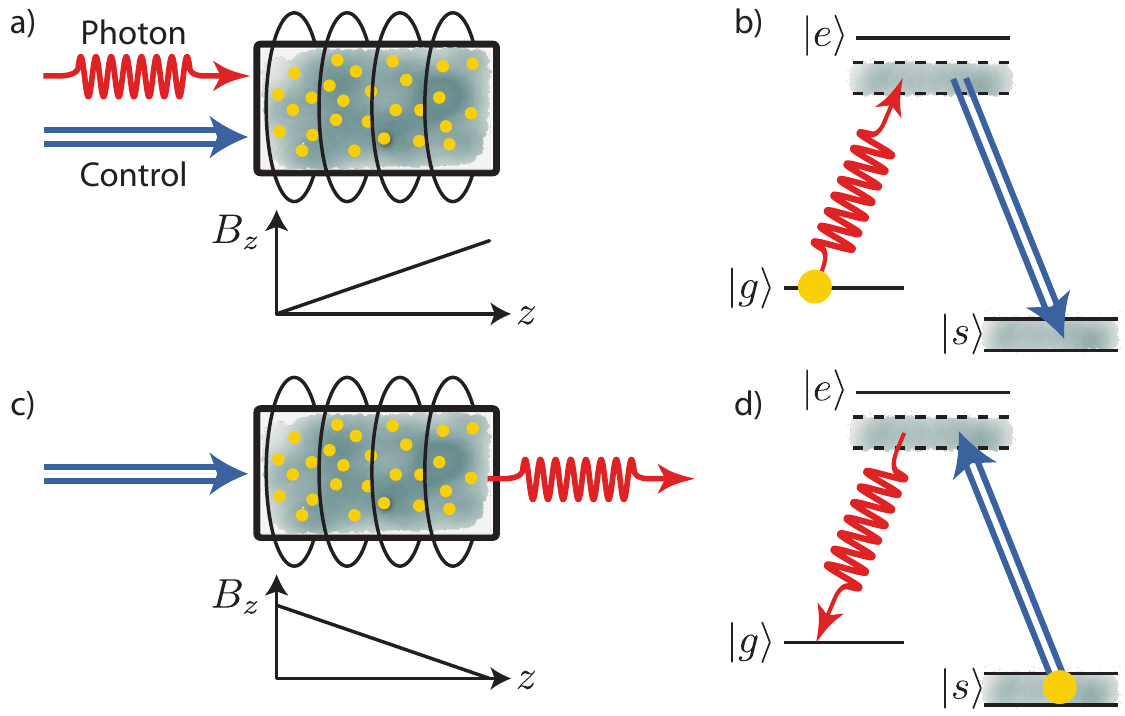}
\caption{$\Lambda$-GEM scheme. a) and b) The $g$-$s$ spin transition of an atomic ensemble is inhomogeneously broadened along the $z$ direction by a magnetic field gradient through the Zeeman effect. The effect of the broadening is illustrated on the energy structure where the $\ket{s}$ state is broadened into a greyed region. An off-resonant photon is sent into the ensemble along with a control field. Absorption of the photon through a Raman interaction creates a spin collective state that starts dephasing right after absorption. c) and d) After reversing the field gradient, the collective dipole moment of the ensemble eventually rephases and can be read out optically with the help of the control beam.}%
\label{fig:GEM}
\end{center}
\end{figure}

Important progress has also been realized towards quantum storage
using hot atomic vapours, whose room-temperature operation constitute an important advantage when considering a large-scale implementation of a quantum repeater. The most important advances have been done
using off-resonant interactions. Several experiments demonstrating
gradient-echo spin-wave memories ($\Lambda$-GEM) have been
reported~\cite{Hetet2008a}. The general idea is to induce a controlled broadening of
the spin transition memory before mapping the light field onto
an atomic spin wave using an off-resonant stimulated Raman
interaction; see Fig.~\ref{fig:GEM}. After the mapping, the spin wave dephases due
to the inhomogeneous spin broadening. After a user-controlled
storage time, the inhomogeneous broadening is reversed, which then
reverses the dephasing of the spin wave. When the dipoles are in
phase again, the spin wave can be read out optically by a strong control
pulse. This scheme has been used to demonstrate a coherent pulse
sequencer for bright pulses in a Rb vapour, where the stored pulses
can be read out in arbitrary order~\cite{Hosseini2009a}. More
recently, it was used to demonstrate a high efficiency coherent
memory in a Rb vapour (also in the classical regime), with a storage
and retrieval efficiency of 89\%, the highest to date for any
system~\cite{Hosseini2011a}. While the scheme is in principle
directly extendable to the quantum regime, the technical noise due
to the strong control pulses needs to be suppressed when working at
the single photon level. This was subsequently achieved as reported in~\cite{Hosseini2011b}, where the authors demonstrate an
efficient and low-noise quantum memory for weak coherent states.
Using a homodyne detection scheme, they show a recall fidelity of
up to 98\% for coherent pulses containing around 1~photon. The
$\Lambda$-GEM scheme has so far not been used to store non-classical states of light.

Another promising Raman scheme has been proposed and demonstrated~\cite{Reim2010a}. It has the ability to store broadband, sub-nanosecond photons. This scheme is also well-suited for linear-optical quantum computing; it is covered in section~\ref{section-LOQC}. 


\subsubsection{Rare-earth doped solids} \label{section:reic}
Finally, we review the recent progress of quantum memories based on rare-earth doped crystals. These solid-state memories take advantage of the long optical and spin coherence times of impurity ions in a crystalline host cooled to cryogenic temperatures. They typically have inhomogeneous broadenings ranging from of a few hundreds of MHz up to hundreds of GHz, which makes them particularly well suited for storing broadband photons with a large multimode capacity. They were reviewed in 2010~\cite{Tittel2010a}, but the field evolved tremendously since then, and several achievements relevant for quantum repeaters have meanwhile been demonstrated. 

One of these achievements is high efficiency. In 2010, a quantum memory for weak
coherent states was demonstrated in Pr$^{3+}$:Y$_2$SiO$_5$~\cite{Hedges2010a} using a two-level gradient-echo quantum memory scheme, which is similar in spirit to the $\Lambda$-GEM discussed in Fig.~\ref{fig:GEM}, but operates on the optical transition only. Storage and retrieval efficiencies of up to 69\% were obtained with a storage time of 1.3~$\mu$s. This is the highest reported efficiency in solid-state quantum memories so far.

Another achievement is long storage time. Very recently, the group of Halfmann has reported an EIT experiment in Pr:YSO crystal, with the impressive storage time of more than 1 minute. The experiment uses external magnetic fields and dynamical decoupling techniques. It pushes the coherence towards the fundamental limit of population decay.

\begin{figure}[!t]
\begin{center}
\includegraphics[scale=0.9]{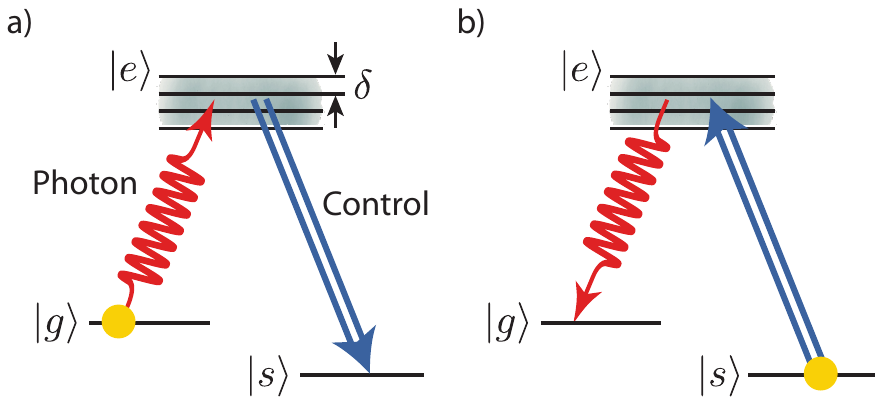}
\caption{Atomic frequency comb storage protocol. a) All atoms are initially in the ground state $\ket{g}$. The inhomogeneously broadened optical transition $g$-$e$ is then shaped into a comb structure with a periodic frequency spacing $\delta$. A single photon whose spectrum covers a few peaks of the comb is then absorbed on the optical transition. The collective dipole moment dephases, but thanks to the periodicity of the comb structure, it rephases automatically at a time $t_e = 1/\delta$. If no control pulse is applied between absorption and rephasing, the large collective dipole moment at time $t_e$ then leads to a re-emission of the stored photon. This yields a quantum memory with a preprogrammed storage time, which we refer to as a \emph{two-level AFC scheme}. The complete AFC scheme with on-demand retrieval is obtained by applying a control pulse to transfer the optical collective state onto a spin wave at time $t_c < t_e$. Provided the $\ket{s}$ level has no inhomogeneous broadening, the dephasing is stopped for a user-programmable time $T$. b) To obtain re-emission, another control pulse is used to transfer the coherence back to the optical transition. Evolution of the collective dipole then resumes until rephasing happens at a time $t_e+T$, which leads to re-emission.}%
\label{fig:reic}
\end{center}
\end{figure}

Storage of light at the single-photon level is another important achievement. In the footsteps of the first storage of a weak coherent state in a rare-earth-ion doped crystal in 2008~\cite{Riedmatten2008a}, storage and recall of weak coherent pulses at the single-photon level was  demonstrated in Pr$^{3+}$:Y$_2$SiO$_5$ with an improved efficiency of 25\%~\cite{Sabooni2010a}. These experiments used the atomic frequency comb (AFC) storage protocol, described in Fig.~\ref{fig:reic}. The efficiency of the AFC protocol is in part limited by the low optical depth of some specific crystals~\cite{Afzelius2009a}. One approach to overcome this limitation is to embed the crystal in an impedance-matched optical cavity~\cite{Afzelius2010b,Moiseev2010a}. This proposal was successfully demonstrated using a weakly absorbing Pr$^{3+}$:Y$_2$SiO$_5$ crystal to store light pulses with a 56\% efficiency during 1~$\mu$s~\cite{Sabooni2013a}. 

The two-level version of the AFC protocol has rapidly proven fruitful to store broadband light at the single-photon level. This paved the way for experiments involving the storage of non-classical light generated from spontaneous downconversion sources, thereby realizing some of the key features of the proposal of Ref.~\cite{Simon2007a} to combine photon-pair sources and multimode memories to achieve a high-rate quantum repeater with transmission over optical fibres~\cite{Sangouard2011a}. In 2011, two groups demonstrated the storage of an entangled photon in a crystal. On the one hand, an energy-time entangled photon with a 43~MHz bandwidth was stored in Nd$^{3+}$:Y$_2$SiO$_5$ with an efficiency of 20\% and a storage time of 25~ns~\cite{Clausen2011a}. The stored energy-time entanglement between the crystal and another photon at 1338~nm was revealed through the violation of a Bell inequality. On the other hand, another experiment reported the storage of an entangled time-bin qubit in a thulium-doped lithium niobate waveguide for 7~ns~\cite{Saglamyurek2011a}. The 5~GHz storage bandwidth of this memory highlighted its high broadband capacity. 
Pure time-bin qubits encoded in heralded single photons have also been stored in the same waveguide~\cite{Saglamyurek2012a}, and two-photon interference and Bell-state measurements between two weak coherent states stored in two separate waveguides was also demonstrated~\cite{Jin2013a}. Another experiment demonstrated heralded entanglement between two separate crystals by mapping a delocalized single-photon entangled state in a out of the crystals~\cite{Usmani2012a}. 

Most of the experiments with the AFC protocol involved the storage of quantum information encoded in the time degree of freedom. Three experiments have recently extended the storage capabilities of rare-earth doped solids to polarization qubits \cite{Clausen2012b,Gundogan2012a,Zhou2012a} by using two crystals arranged in a way that effectively yields polarization-independent storage efficiency and fidelity, despite of the fact that each crystal's absorption coefficient is strongly dependent on the polarization of the light to be stored. 

The AFC experiments described so far were done with storage in excited state only, leading to pre-determined and short storage times. On-demand read-out and long storage times can be obtained by transferring the excitation to a  spin-state level, which requires materials with at least three ground state levels. In that case, the memory bandwidth is limited by the spacing between hyperfine levels. The first demonstration of this complete AFC scheme came in 2010~\cite{Afzelius2010a} (see Fig.~\ref{fig:reic}). It has recently been shown to preserve the coherence of time-bin pulses, and up to 5 temporal modes have been stored~\cite{Gundogan2013a}. Storage of light pulses containing an average of 2.5~photons level was also demonstrated~\cite{Timoney2013a}. 

The experiments mentioned above describe quantum memories in which an optical
input field is stored and retrieved. Some proposals have been put
forward to use rare-earth doped crystals as sources of
quantum light with embedded memory based on
rephased spontaneous emission (RASE)~\cite{Ledingham2010a} or a
combination of DLCZ like source and AFC memories~\cite{Sekatski2011a}. The RASE protocol has been recently
demonstrated by two groups, in a thulium-doped crystal
\cite{Ledingham2012a} and in a Praseodymium-doped crystal
\cite{Beavan2012a}. In Ref.~\cite{Ledingham2012a}, homodyne
detection was used, and time separated non-classical correlations
between the spontaneously emitted optical pulse (coming first), and the rephased
one (coming second), were observed. In Ref.~\cite{Beavan2012a}, the authors reported the
observation of photon-echo rephasing of spontaneous emission,
using photon counting techniques. Strong classical correlation between
spontaneous emission was observed. This paves the way for heralding entanglement between 
quantum memories through the emission of photons whose spectra is directly tailored by the crystals themselves, and would require little or no spectral filtering. 

Their ease of use and the potential large bandwidth of rare-earth-ion doped crystals constitute important advantages over narrowband atomic ensembles. The technical challenges that emerge from exploiting this large bandwidth are, one the one hand, the need for involved optical pumping (whose efficiency remains limited in some materials) to initialize the memory and to transfer the population from optical to spin transitions, and on the other hand the need for precise manipulation of the spin wave with radio-frequency fields to overcome dephasing (i.e.~dynamical decoupling). This was studied in~\cite{Heshami2011a} which showed that this within the reach of realistic levels of precision.

\begin{figure}[!t]
\begin{center}
\includegraphics[scale=0.9]{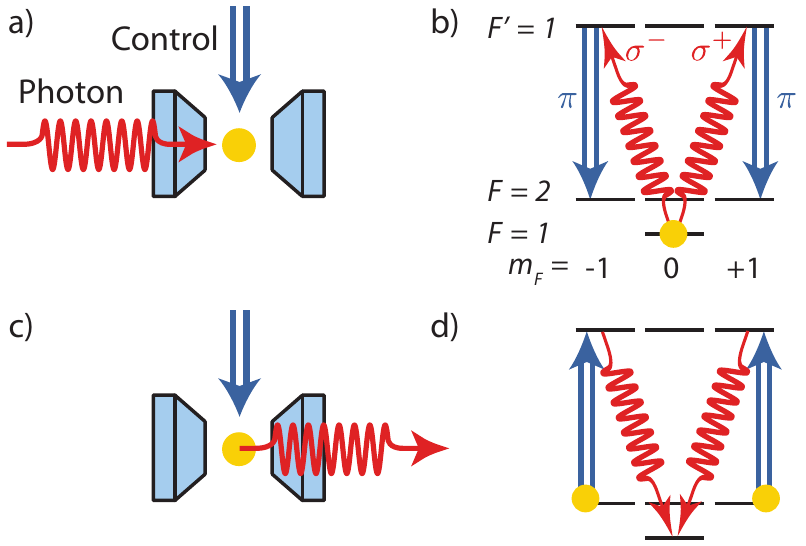}
\caption{Storage in a single atom trapped in a high-finesse optical cavity~\cite{Specht2011a}. a) A photon with arbitrary polarization is resonant with the $\ket{F=1}$-$\ket{F'=1}$ transition of $^{87}$Rb. Absorption of the photon maps its polarization state onto a superposition of the $\ket{F=2,m_F=\pm1}$ states through interaction with a $\pi$-polarized control field. b) Re-emission is triggered by applying the control field again.}%
\label{fig:single}
\end{center}
\end{figure}

\subsubsection{Single quantum systems} \label{section:singlequantumsystems}
In the past few years, experimental control of single quantum systems has progressed to the point where 
elementary building blocks of quantum networks could be demonstrated. We first review some of the impressive progress realized with single atoms embedded in high-finesse cavities. A
quantum memory for weak coherent state at the single-photon level
has been demonstrated using a single Rb atom
by mapping arbitrary polarization states of light onto atomic spin
superpositions using strong control pulses~\cite{Specht2011a}; see Fig.~\ref{fig:single}. A storage and
retrieval efficiency of 9\%, an average fidelity of 93\%
and a storage time of 180~$\mu$s was obtained. The same group
demonstrated later an elementary quantum network using two single-atom 
quantum memories in separate atom traps distant by 21~m
\cite{Ritter2012a}.  They achieved quantum state transfer and
entanglement between the two atoms by transmission of a single
photon. The quantum state of the atoms is measured by transferring
the atomic state to light polarization and by measuring the resulting
photons. Heralded entanglement between two single atoms sitting in dipole traps and separated by 20~m was also
demonstrated in 2012~\cite{Hofmann2012a}. This work is discussed in section~\ref{section-foundations}.

The impressive control of single atoms in cavities makes them versatile building blocks to distribute and establish entanglement between nodes in a small-scale network. However, their limited multimode capacity imposes a hard limit to the rate at which entanglement can be distributed in a quantum repeater. Nonetheless, this is alleviated by the potential capability to implement the Bell state measurements deterministically. 

Entanglement between single quantum systems is not restricted to single atoms and ions. It has now also been demonstrated with single spins in a solid-state environment. In 2010, entanglement between the spin of a single nitrogen-vacancy centre (NVC) in diamond and an optical photon was demonstrated~\cite{Togan2010a}. Improvement in the control of the emitted photon's spectrum eventually led to quantum interference between photons emitted by two NVC~\cite{Bernien2012a,Sipahigil2012a} and, subsequently, to the demonstration of heralded entanglement between two NVC separated by 3~meters~\cite{Bernien2013a}. Similarly, entanglement between a single quantum dot and an optical photon was recently demonstrated~\cite{DeGreve2012a,Gao2012a}.

\section{Linear-optical quantum computation} \label{section-LOQC}
Linear-optical quantum computing (LOQC) is based on probabilistic quantum gates using linear optics and single-photon sources~\cite{Knill2001a,Kok2007a}. In order to make such schemes scalable beyond a few qubits and a few gates one would need to employ quantum memories in order to synchronize probabilistic and independent events such as the creation of photon pairs subjected to transmission loss. LOQC also requires many efficient, on-demand single-photon sources. Quantum memories could in principle also enhance the photon generation rate of sources used in multiphoton experiments~\cite{Yao2012a,Prevedel2009a,Wieczorek2009a}. 

\begin{figure}[!t]
\begin{center}
\includegraphics[scale=0.9]{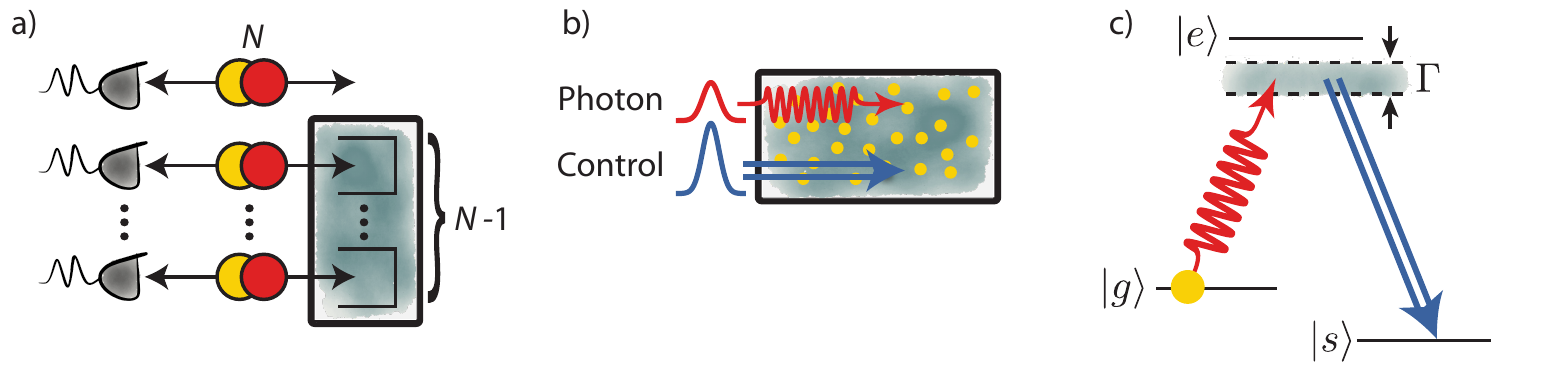}
\caption{Linear-optical quantum computation with quantum memories. a) The scenario analyzed in~\cite{Nunn2013a} consists of $N$ sources of photon pairs continuously triggered until all of them have heralded single photons. Buffering $N-1$ of these photons in quantum memories, to store them as offline resources until all other sources heralded photons, yields a significant reduction in the average time required to successfully herald the $N$-photon state. This remains true even if the memories are less than 100\% efficient. b) and c) Broadband Raman quantum memory scheme. All atoms are initially in the $\ket{g}$ state. A photon off-resonant with the $g$-$e$ transition is sent along with a strong, short control pulse that is off-resonant with the $s$-$e$ transition. The large bandwidth $\Gamma$ of the control pulse (illustrated by the greyed region underneath the $\ket{e}$ level) transfers the stored excitation to a spin wave over a spectral width that can be as large as $\Gamma$, thus allowing to store broadband photons. Retrieval is obtained using another broadband control field that reverses the process.}%
\label{fig:LOQC}
\end{center}
\end{figure}

The desired properties of quantum memories for LOQC and other multi-photon experiments are somewhat different than those required for quantum repeaters. While high efficiency and high fidelities are also required, there are less constraints on the memory time. Quantum repeaters are meant to work over distances of several hundreds of kilometers, while LOQC aims at small-scale devices using, for example, integrated reconfigurable optical circuits~\cite{Shadbolt2012a}. For quantum repeaters it has been shown that multimode capacity is crucial for long-term scalability, but absolute memory bandwidth is less important since the basic rate is determined by the communication time between network nodes. For LOQC, the absorption bandwidth $\Gamma$ of quantum memories is crucial because it fixes the smallest temporal width of the photons, and thus directly affects the overall clock rate of the linear-optical circuit. An important parameter for LOQC quantum memories is thus the time-bandwidth product (TBP), which relates the memory bandwidth $\Gamma$ to the memory storage time $T$. This parameter is related to how many times a probabilistic event can be repeated (until success is achieved) while other photons are buffered in a quantum memory. It should be emphasized that quantum memories for repeaters also need high TBP, the specificity of LOQC is the additional need for high bandwidth in order not to limit achievable rates. The necessity of quantum memories for LOQC and multiphoton experiments was recognized early on, yet only a few studies have been published on the benefit of quantum memories in such schemes~\cite{Barrett2010a,Nunn2013a}. Recently, the performance of a multi-photon source based on multi single-photon sources and quantum memories was analyzed~\cite{Nunn2013a}; see Fig.~\ref{fig:LOQC}-a. It is shown that even inefficient memories can significantly enhance the output rate and that the relevant factor in the enhancement is given by $\eta \Gamma T$, where $\eta$ is the storage and retrieval efficiency of each memory. There is clearly room for much more theoretical and experimental work in this direction. Below we review some experimental work on quantum memories that is relevant in the context of synchronization and clock-rate enhancement in multi-photon experiments.

An intrinsically broadband quantum memory scheme based on a far-detuned off-resonant Raman interaction in warm alkali vapours has been proposed by in Ref.~\cite{Nunn2007a}, and has the particularity that it can store sub-nanosecond photons; see Fig.~\ref{fig:LOQC}-b-c. The first experimental demonstration in 2010 used a caesium-vapour cell heated slightly above room temperature~\cite{Reim2010a}. The two-photon transition was detuned by 9 GHz from the one-photon resonance, allowing storage of a 300-ps long pulses of strong light for 12.5~ns. This demonstration in the classical regime was quickly followed by a proof-of-principle experiment at the single-photon level through the storage of a coherent state with a 1.6 average number of photons in the stored pulse~\cite{Reim2011a}. The storage time was increased to 4~$\mu$s, yielding a TPB of 2500, which is one of the highest achieved so far. The measured unconditional noise floor produced by the strong Raman control fields was 0.25 photons per pulse.
Since the storage efficiency was at most 30\%, this resulted in a signal-to-noise ratio of around one. Recently, the same group also showed that this kind of memory can serve as temporal mode selective beam splitter, by using multiple control pulses~\cite{Reim2012a}. In 2012, it was also shown that this Raman memory can store the polarization state of light~\cite{England2012a} using the dual-rail technique. The quality of the storage was tested by quantum process tomography. The measured process fidelity $F$ stayed roughly constant for storage times between 12.5 ns ($F$=97$\%$) and 1.5~$\mu$s ($F=86\%$). The optical pulses defining the polarization qubit contained 1000-10000 photons. The impact of the unconditional noise floor for true single-photon storage has yet to be evaluated and measured. While this Raman scheme memory scheme has the potential to operate in the quantum regime, further noise reduction is necessary for implementation in LOQC experiments.

It is also worth mentioning the recent demonstration of active feedforward in a simple one-way quantum computing experiment based on a narrow-band laser-cooled atomic ensemble memory~\cite{Xu2012a}. Entanglement rate enhancement using laser-cooled quantum memories for quantum repeaters were also demonstrated~\cite{Felinto2006a,Yuan2007a}. In this context, Ref.~\cite{Barrett2010a} proposed a scalable quantum computing scheme using similar experimental resources. 

\section{Quantum metrology and magnetometry} \label{section-metrology}
Entanglement between the atoms of an ensemble can improve dramatically the precision of certain measurements, which is especially interesting for the field of metrology~\cite{Giovannetti2011a}. 

We first describe the idea of the Ramsey interferometer, which is central to this discussion. We consider a two-level atom (here described as a spin-$\fud$ particle) that is optically prepared in the superposition state $\frac{1}{\sqrt{2}}(\sd + \su)$, and that evolves to $\frac{1}{\sqrt{2}}(\sd + \e^{i\varphi}\su)$ after a time $t$, where $\varphi = (E_{\uparrow}-E_{\downarrow}) t/\hbar$. The phase $\varphi$ can be estimated using Ramsey interferometry, i.e.~by mapping the phase onto probabilities of occupying the ground and excited states, which are then measured. Implementing this on $N$ independent and identically prepared atoms of an ensemble allows one to estimate the phase with a relative uncertainty $\delta \varphi = N^{-1/2}$; this is known as the standard quantum limit (SQL). This scaling originates from \emph{projection noise}, i.e.~from the probabilistic nature of projective measurements on single quantum systems.

\begin{figure}[!t]
\begin{center}
\includegraphics[scale=0.8]{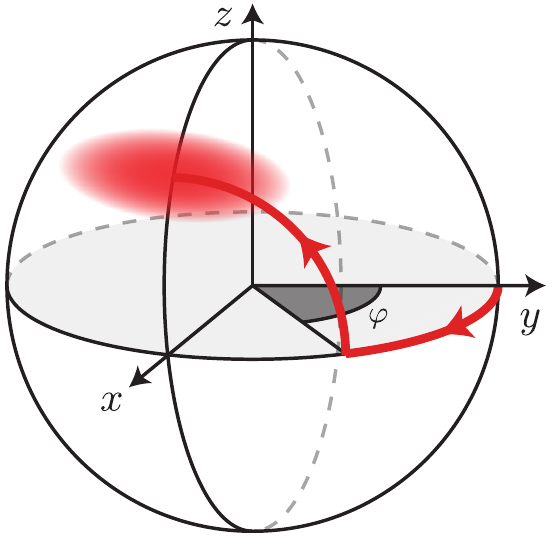}
\caption{Quantum metrology with spin-squeezed atomic states. This drawing represents the evolution of the collective pseudo spin of a suitably prepared atomic ensemble in a Ramsey interferometer. The ensemble is optically prepared such that its pseudo spin initially points towards +$y$. Free evolution then results in a constant rotation around $z$ by a phase $\varphi$. Another pulse finally maps the state on the $xz$ circle, such that a projective measurement of the $z$ component of the pseudo spin allows to estimate the value of $\varphi$. The red oval represents the spread of the final position of a suitably prepared spin-squeezed state, the most probable value being at the centre and the least probables ones towards the edges. The fact that the oval is elongated along the $y$ axis means that the fluctuations $\Delta S_z$ are reduced, thereby increasing the precision in estimating $\varphi$ beyond the standard quantum limit.}%
\label{fig:metrology}
\end{center}
\end{figure}

Interestingly, the SQL can be surpassed by dividing the $N$ atoms into $m$ subgroups of $n = N/m$ atoms prepared in carefully chosen entangled states. The optimal choice of state yields the uncertainty $\delta \varphi = (nN)^{-1/2}$, which is known as the Heisenberg limit. In particular, $\delta \varphi = N^{-1}$ when all the atoms are entangled, which is smaller the SQL by a factor of $\sqrt{N}$. In atomic ensembles, entanglement in the form of spin-squeezed states (SSS) can also surpass the SQL; see Fig.~\ref{fig:metrology}. These states can be given a simple interpretation using the pseudo spin of the ensemble, $\vect{S} = \sum_{i=1}^N \vect{s}_i$, whose length lies in the range $0\leq S \leq N/2$ and that can be represented as a vector on a Bloch sphere. In this representation, the component $S_z$ is proportional to the population difference and is related to the phase $\varphi$ by a Ramsey sequence as described  above. In a SSS, the fluctuations $\Delta S_z$ are reduced below the SQL, thanks to the multi-atom entanglement. 
In practice, the necessary steps to prepare a SSS can, however, decrease the length of the pseudo-spin,
which in return reduces the fluctuation $\Delta S_z$. One must then consider the \emph{metrologically} relevant squeezing by comparing the resulting fluctuations with those that could have been obtained without the squeezing.

Atomic ensemble are thus excellent candidates to realize SSS with a very large number of atoms, and their demonstration has been an active research topic for the last few years. SSS in atomic ensembles was demonstrated in Bose-Einstein condensates by exploiting collisional interactions between particles trapped in an optical lattice~\cite{Esteve2008a} or using an atom chip~\cite{Riedel2010a}. SSS can also be generated by optical means using optical quantum non-demolition (QND) measurement of the collective atomic spin of the ensemble~\cite{Kuzmich2000a}. SSS generated in this way can, in a way, be seen as optically-programmed quantum memories storing multi-atom entangled states. This does not correspond to the commonly understood notion of an optical quantum memory. Nevertheless, atomic SSS are closely connected to optical quantum memories, and reviewing some of the recent experimental achievements in SSS is relevant to quantum memories in general. 

One very important application of SSS is atomic clocks~\cite{Santarelli1999a}, whose precision improves as the fluctuations $\Delta S_z$ decrease. The very first demonstration of a SSS on a clock transition came from two independent experiments. In 2009, the demonstration of 3.4~dB of metrological squeezing of $10^5$ caesium atoms held in a dipole trap using an optical quantum non-demolition (QND) measurement was reported~\cite{Appel2009a}. In 2010, a similar result on $5\times 10^4$ rubidium atoms ($^{87}$Rb) trapped inside an optical cavity was demonstrated~\cite{Schleier2010a}. The role of the cavity was to strongly couple the ensemble to a resonator mode to perform the QND measurement. The enhancement allowed to measure $S_z$, and its fluctuations, through the frequency shift of the resonator mode.

Strong ensemble-light coupling in a cavity was also used to generate a SSS on a clock transition without resorting to QND measurements, which alleviates all detrimental detector imperfections. Specifically, it was recently demonstrated that by carefully positioning a cavity mode with respect to a canonical microwave clock transition of rubidium, the frequency shift of the mode induced by the atoms can be made to depend on $S_z$, whose value can thus be probed using a laser~\cite{Leroux2010a}. The coupling between the atoms and the probe laser then generates the entanglement, which led to the demonstration of more than 5~dB of metrological squeezing on $5\times 10^4$ rubidium atoms. 

Another application is atomic magnetometry, i.e.~the measurement of magnetic fields using atomic ensembles~\cite{Budker2007a}. Akin to atomic clocks, the sensitivity of an atomic magnetometer depends on the precision with which one can measure small rotations of the magnetization of a spin-polarized ensemble when it is subjected to an external magnetic field. This requires overcoming several obstacles, notably the absolute calibration of the noise of an optical QND measurement of the magnetization of the ensemble. This was achieved for the first time in 2010 on a magnetically-sensitive transition of a cold $^{87}$Rb ensemble, which paved the way to the demonstration of sub-projection-noise sensitivity to broadband magnetic fields~\cite{Koschorreck2010a}. Further work then lead to the demonstration of spin squeezing and multi-atom entanglement using a magnetically-sensitive transition~\cite{Sewell2012a}. Another approach focused on the use of two suitably prepared ensembles of atomic caesium vapour to reach projecting-noise-level sensitivity to radio-frequency magnetic fields~\cite{Wasilewski2010a}. Furthermore, it was shown that entanglement between the two ensembles could yield a sensitivity comparable to state-of-the-art atomic magnetometers operating with a much larger number of atoms~\cite{Wasilewski2010a}.

There are also proposals for enhanced metrology that do not rely on spin squeezing. Ref.~\cite{Brunner2011a} proposed a heralded amplification scheme for small rotations of the collective spin of an atomic ensemble. The scheme makes use of basic operations that are essential in the context of quantum memories as well, namely the mapping of quantum states between light and collective atomic excitations. The amplification factor is approximately $p^{-\fud}$, where $p$ is the success probability. This technique therefore helps in situations in which the measurement precision is limited not purely by statistics. The ability to amplify small rotations of collective spins might be valuable in the context of atomic clocks, or for weak-field magnetometry. 

Another way of enhancing measurement precision is via the use of so-called NOON states in interferometry, i.e.~states of the form $\ket{N,0}+\ket{0,N}$~\cite{Giovannetti2011a}. Ref.~\cite{Wei2011a} proposed a method for creating such states, where the entangled subsystems are collective atomic excitations, which can be created from photons and converted back into photons using standard quantum memory techniques. The method uses the phenomenon of Rydberg blockade, which is due to the strong interactions between atoms in certain highly excited states (Rydberg states). The proposed technique may be useful for the detection of small displacements or for inertial sensors based on atomic clouds.

\section{Single-photon detection with quantum memories} \label{section-detection}

Many applications of quantum optics involving quantum memories share a common feature: the number of required photons increases steadily as  experiments progress. This is especially true for quantum repeaters and linear-optical quantum computing. Therefore, the use of highly-efficient, noiseless single-photon detectors would facilitate their developments. This can be achieved with a quantum memory, as proposed in 2002~\cite{Imamoglu2002a,James2002a}, through the use of an ensemble of atoms to convert a single photon into fluorescence that is easily detectable. The method not only promises to detect single photons with very high efficiencies and low noise, but also features a photon number resolution as the intensity of the fluorescence increases with the number of detected photons, as we shall see below. 

\begin{figure}[!t]
\begin{center}
\includegraphics[scale=0.8]{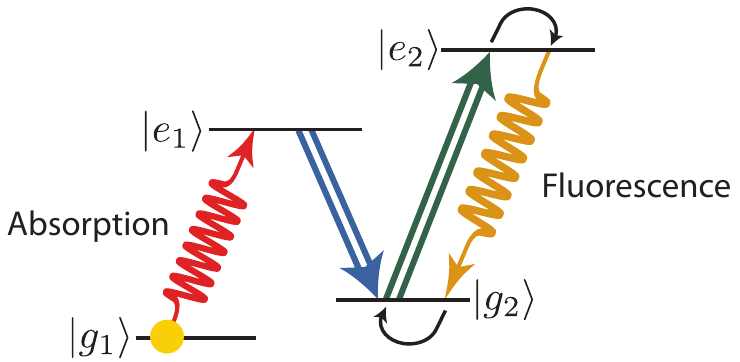}
\caption{Single-photon detector using an optical quantum memory and fluorescence detection. All atoms are initially in the ground state $\ket{g_1}$. A single photon resonant with the $g_1$-$e_1$ transition can then be absorbed. The excitation is transferred to a collective spin wave between states $\ket{g_1}$ and $\ket{g_2}$ using light resonant with $g_2$-$e_2$ transition. Another light beam coupling $\ket{g_2}$ to $\ket{e_2}$ exclusively (using selection rules) then excites the atom. The latter spontaneously decays back to $\ket{g_2}$, only to be excited again, and so on. The resulting fluorescence is detected, and its intensity is proportional to the number of photons detected.}%
\label{fig:detection}
\end{center}
\end{figure}

The principle of the proposals presented in~\cite{Imamoglu2002a} combines two key ingredients: an optical quantum memory and fluorescence detection originally developed for experiments with ion traps; see Fig.~\ref{fig:detection}. Specifically, a photonic state associated with a propagating light pulse is first mapped onto collective spin excitations, similarly to what is realized during the storage step of any quantum memory protocol. The resulting number of spin excitations is subsequently measured efficiently by using state-selective fluorescence measurements. The latter proceeds by coupling exclusively the excited spin state to an excited fluorescing state that decays predominantly by spontaneous emission back to the same excited spin state (a \emph{cycling} transition). Monitoring the intensity of the fluorescence allows one to know the number of spins that have been excited through the absorption process, i.e.~the number of photons contained in the input state.

Although the proposal dates back to 2002, it has not been experimentally implemented yet, since finding a suitable system is quite challenging. Refs.~\cite{Imamoglu2002a,James2002a} chose hyperfine states of the ground electronic level of an alkali atom, like caesium, for the spin states as they exhibit long coherence times. However, this advantage also has drawbacks. To guarantee the absorption of the input pulse with an efficiency approaching unity, about a million of atoms is required. However, the probability for having an unwanted off-resonant excitation of an atom in the ground spin state during the detection stage (which is given by the square of the ratio between the linewidth of the excited fluorescing state and the hyperfine splitting) is estimated to be on the order of $10^{-6}$ in this system. This means that the noise is comparable to the signal from a few photon input pulse. The fluorescence based measurement also imposes strong constrains on the efficiency of the initialization of the atoms, since any excitation left in the excited spin state can contribute to the fluorescence at the detection stage. Furthermore, during the time that is needed to collect the fluorescence detection, premature loss of atoms from the trap can occur due to heating or light-assisted collisions, thus reducing the readout fidelity.

The above mentioned problems may be circumvented by using an ensemble of ions embedded in a moderate finesse cavity. The analysis reported in Ref.~\cite{Clausen2012a} showed that a crystal realized with 1500 ions and a cavity with a finesse of 3000 would allow essentially noiseless detection with an efficiency larger than 90\%. On the downside, this analysis shows that an ion-based detector would be rather slow in term of repetition rates (about 3 kHz), limited by the time needed to collect enough fluorescence detection and by cooling of the ions between trials. It may nevertheless find applications in long-distance quantum communication where the repetition rates are limited to the time a photon takes to travel over tens or even hundreds of kilometres. 

\section{Foundations of quantum mechanics} \label{section-foundations}
\begin{figure}[!t]
\begin{center}
\includegraphics[scale=0.9]{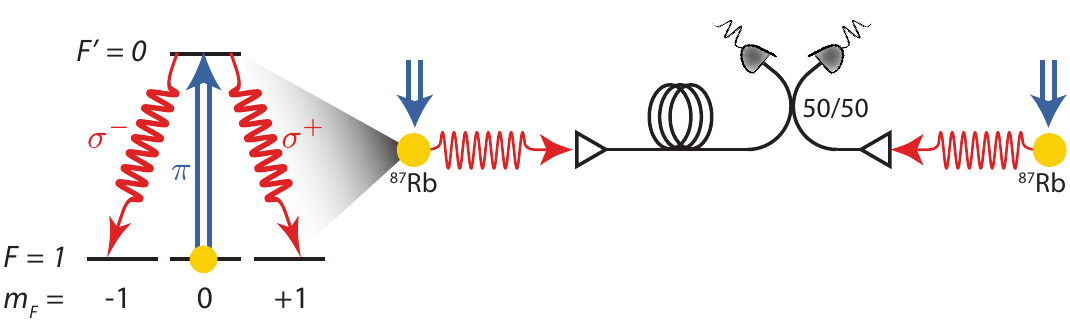}
\caption{Heralded entanglement between remote trapped ions~\cite{Hofmann2012a}. Two 20-metre distant rubidium-87 atoms are loaded in dipole traps. Each atom is excited with a $\pi$-polarized pulsed from the $5^2\text{S}_{1/2}\ket{F=1,m_F=0}$ state to the $5^2\text{P}_{3/2}\ket{F'=0,m_F'=0}$ state. Spontaneous decay occurs to an even superposition of the $\ket{F=1,m_F=\pm1}\otimes\ket{\sigma^{\pm}}$ states (where $\ket{\sigma^{\pm}}$ represents right and left-circular polarization states of the emitted photon), thereby creating a photon whose polarization is entangled with the atom. The two photons emitted from both traps are sent into a 50/50 beamsplitter which realizes a Bell state measurement, hence heralding entanglement between the two remote atoms. Violation of a Bell inequality was observed after measuring the state of the atoms directly. 
}%
\label{fig:swapping}
\end{center}
\end{figure}

Probing the foundations of quantum mechanics has been a long-standing goal since the pioneering work of John Bell on nonlocality. To date, no experiment has closed the detection and locality loopholes simultaneously. Quantum memories based on single trapped atoms (or ions) offer a promising approach towards this goal. On the one hand, the quantum state of one ion can be read with close to unit efficiency using a cycling transition (see section~\ref{section-detection}). On the other hand, two remote ions can be entangled by first creating light-matter entanglement between an ion and a photon, followed by a Bell state measurement on the two photons emitted by the ions. The ions are left entangled in a heralded fashion, and a careful time ordering of all the events could close both loopholes~\cite{Simon2003a}. Significant progress towards the realization of this proposal was reported in Ref.~\cite{Hofmann2012a} in which two single rubidium-87 atoms separated by 20 meters were successfully entangled in a heralded fashion; see Fig.~\ref{fig:swapping}. By increasing the distance between the atoms and implementing the recently demonstrated fast atomic detection technique~\cite{Henkel2010a}, this experiment may soon allow closing both loopholes.


Trapped ions have also been shown to be an enabling technology for device-independent quantum information processing. In 2010, it was shown that the use the nonlocal correlations of quantum mechanics allows one to generate certified random numbers~\cite{Pironio2010a}. This form of certification does not depend on any assumption about the internal workings of the device used to generate the numbers, and is a truly fascinating application of nonlocality. The experimental demonstration of this idea is challenging as it requires closing the detection loophole. This was accomplished by using two trapped Ytterbium-171 ions entangled in a heralded fashion, and led to the generation of more than 40 random bits certified by Bell's theorem~\cite{Pironio2010a}.

\section{Quantum information processing with optical quantum memories} \label{section-outlook}

One of the initial goal of this review was to provide motivations for experimentalists to improve the performances of available quantum memories. To do so, we presented various applications, each of them coming with specific requirements. There is another motivation that we did not mention so far, but this review would not be complete without a few words about the potential usefulness of quantum memories for processing the quantum states of individual photons. 

It is known for a along time that a dielectric medium can be used to manipulate the properties of light pulses. 
However, the available nonlinearities are weak. Together with the optical absorption, they limit the extent of possible control and prevent operations on single photons. The utility of quantum memories would be greatly enhanced if they could provide a solution to manipulate light pulses down to the single photon level. 

First, it is worth reminding that a quantum memory can obviously be used to manipulate the quantum state of single photons. When a given photonic qubit is mapped onto a collective spin wave, the latter can be manipulated optically to target a given state of the retrieved photon. Ref.~\cite{Hosseini2009a} provided an elegant illustration by operating on time-bin type encoding with classical light. The storage, based on a two-level gradient echo scheme, allows to store multiple pulses of light within a chosen frequency bandwidth. The authors showed that stored pulses can be recalled in arbitrary order with any chosen delay between each recalled pulse. Moreover, each pulse can be compressed, stretched or split into multiple smaller pulses and recalled in chosen time bins.

The realization of two-photon nonlinear interactions is more challenging. Two pulses propagating through an atomic ensemble can exchange a phase through the optical Kerr effect. The latter is weak in general, but EIT type processes can significantly enhance it so that it may operate on single photons. The key idea is that in an EIT medium, the transparency resonance is accompanied by a very sharp variation of the refractive index with frequency (which, as a result, reduces the group velocity of the light pulse propagating into the medium). Therefore, a slight modification of the energy of the spin transition caused by the interaction of a weak pulse through the Stark effect changes significantly the phase of the re-emitted light pulse. The idea dates back to 1998~\cite{Schmidt1996a}, but an experimental implementation at the single-photon level is still awaited. Several problems need to be solved~\cite{Shapiro2006a,Banacloche2010a}. The main one is the multi-mode character of the light pulses that prevents the realization of high-fidelity two-photon operations. The phase shifts due to the interaction indeed depend on the relative position of the two photons and take different values over the pulses because the photon are not described by point particles but instead as extended wave packets. Potential solutions have been proposed but they lead to more sophisticated gate designs. 
Recently, Ref.~\cite{Hosseini2012a} proposed a solution based on the storage of the light fields using the $\Lambda$-GEM storage protocol (see section~\ref{section:atomicvapors}). They also presented a proof-of-principle experimental demonstration~\cite{Hosseini2012a} with one pulse stored, while the other was freely propagating through the memory medium.

Other proposals use the unique properties of specific atomic systems. For example, Rydberg-blockade-mediated two-photon interaction has been largely investigated over the last years~\cite{Gorshkov2011a}. The underlying principle is that an atom promoted to a Rydberg level shifts the energy levels of nearby atoms, therefore suppressing their excitation. First experiments have been reported recently where the Rydberg blockade in cold atomic gases has been used to extract single photons from a coherent beam~\cite{Dudin2012a,Peyronel2012a,Maxwell2013a}. 

Another example relies on the use of Bose-Einstein condensates (BEC) with long storage times to mediate two-photon interaction. 
A recent work~\cite{Rispe2011a} proposed to store two individual photons in a Bose-Einstein condensate (BEC), and to use the collisional interactions between the atoms in the BEC in order to generate a conditional phase shift. The collisional interactions are quite weak; it is estimated that interaction times of order one second may be required in order to implement a conditional phase shift of $\pi$~\cite{Rispe2011a}. 
However, it has been shown that light can be stored in BECs for several seconds~\cite{Zhang2009a}, so this could be a viable approach towards photon-photon gates in situations where gate times are not the limiting factors, e.g.~for quantum repeaters.

The optical nonlinearity of atomic ensembles can also be exploited for quantum-enhanced metrology. It has been shown by several authors (see~\cite{Giovannetti2011a} for a review) that the uncertainty $\delta \varphi$ on the parameter $\varphi$ to be estimated can in principle scale as $N^{-k}$, where $N$ is the number of photons used and $k$ is the order of the nonlinear light-matter interaction. When photons are constrained to be in a separable state, the optimal scaling is instead $N^{-k-1/2}$. The first experimental demonstration of this scaling was recently obtained in the measurement of the magnetization of an ensemble of $7\times 10^5$ $^{87}$Rb atoms held in an optical dipole trap~\cite{Napolitano2011a}. Light-matter second-order nonlinearity on $10^5$ to $10^7$ photons prepared in a separable state yielded a scaling $N^{-3/2}$, well beyond the $N^{-1}$ Heisenberg limit. 

Many other promising systems such as atoms confined in a hollow-core photonic band gap fibre~\cite{Bajcsy2009a,Venkataraman2013a}, and atoms trapped in the evanescent field around a nano-fibre~\cite{Vetsch2010a}, are very attractive as the confinement provides a strong coupling between few atoms and photons. 

To conclude, we have reviewed several applications of quantum memories, as well as some of the recent experimental progress geared towards the realization of these applications. The different applications come with their own set of requirements for quantum memories, and most of the experimental efforts are aimed at demonstrating the possibility of satisfying one or a subset of these specific requirements. The road towards the development of fully functional quantum memories is clear. On the one hand, integration of all the desired properties in a single system is necessary. On the other hand, the addition of nonlinear processing abilities with quantum memories will undoubtedly prove useful for optical QIP in general. Above all, one fact stands: optical quantum memories are no longer a theoretical wonder. Their prominence is rising, and they are now an essential part in an increasing number of applications of optical quantum information processing. 


\section*{Acknowledgements} We thank Almut Beige, Nuala Timoney and Philipp Treutlein for helpful discussions. 
This work was supported by the EU projects Qscale and ERC-QuLIMA, by the Alberta Innovates Technology Futures, and by the Swiss NCCR QSIT.

\bibliographystyle{tMOP}
\bibliography{biblio-review-JMO}

\end{document}